\documentclass{ifacconf}

\usepackage{natbib}        
\usepackage{amsmath}
\usepackage{amssymb}
\usepackage{graphicx}
\usepackage{siunitx}
\usepackage{booktabs}
\usepackage[dvipsnames]{xcolor}
\usepackage{multirow}
\usepackage{tikz}
\usepackage{graphicx}
\usepackage{scalerel}
\usepackage{soul}

\theoremstyle{break}
\newtheorem{problem}[thm]{\textbf{Problem}\\}

\begin{document}
\begin{frontmatter}
\vspace{-10pt}
\title{ \makebox[0.4\textwidth][c]{Two-wheel-driven~Electric~Superbike~Powertrain~Optimization} }



\author[First]{Adelmo Niccolai} 
\author[Second]{Maurizio Clemente} 
\author[Second]{Theo Hofman}
\author[First]{Niccolò Baldanzini}

\address[First]{Department of Industrial Engineering, University of Florence \\ Via di Santa Marta 3, Florence, 50139, Tuscany, Italy\\e-mails: \{adelmo.niccolai, niccolo.baldanzini\}@unifi.it}
\address[Second]{Eindhoven University of Technology\\ 
	5600 MB Eindhoven, The~Netherlands\\e-mails: \{m.clemente, t.hofman\}@tue.nl}

\begin{abstract}  
In this paper, we propose an optimization framework for the powertrain design of a two-wheel-driven electric superbike, minimizing energy consumption.
Specifically, we jointly optimize the force distribution between the wheels with the gear ratio, and rear motor and battery sizing while explicitly considering vehicle dynamics and performance constraints. 
First, we present an energy consumption model of the vehicle, including a scalable model of the electric machine based on data from the industry, accounting for iron, copper, and mechanical losses.
Then, we analyze the propulsive blending strategy to distribute the required power to the wheels while considering adherence limits.
Finally, we demonstrate the effectiveness of our approach by analyzing the design of a superbike, based on regulatory driving cycles and a custom high-performance circuit by comparing the force distribution approaches.
The results underline the significance of joint optimization of powertrain components and propulsive bias, achieving a reduction of up to 22.36\% in energy consumption for the Sport high-performance driving cycle.
 

\end{abstract}

\begin{keyword}
 E-motorbike, Powertrain Optimization, Optimal Propulsive Blending Strategy, Two-wheel-driven.
\end{keyword}

\end{frontmatter}

\section{Introduction}\label{sec:introduction}
\begin{figure}
    \centering
    \includegraphics[width=0.85\columnwidth]{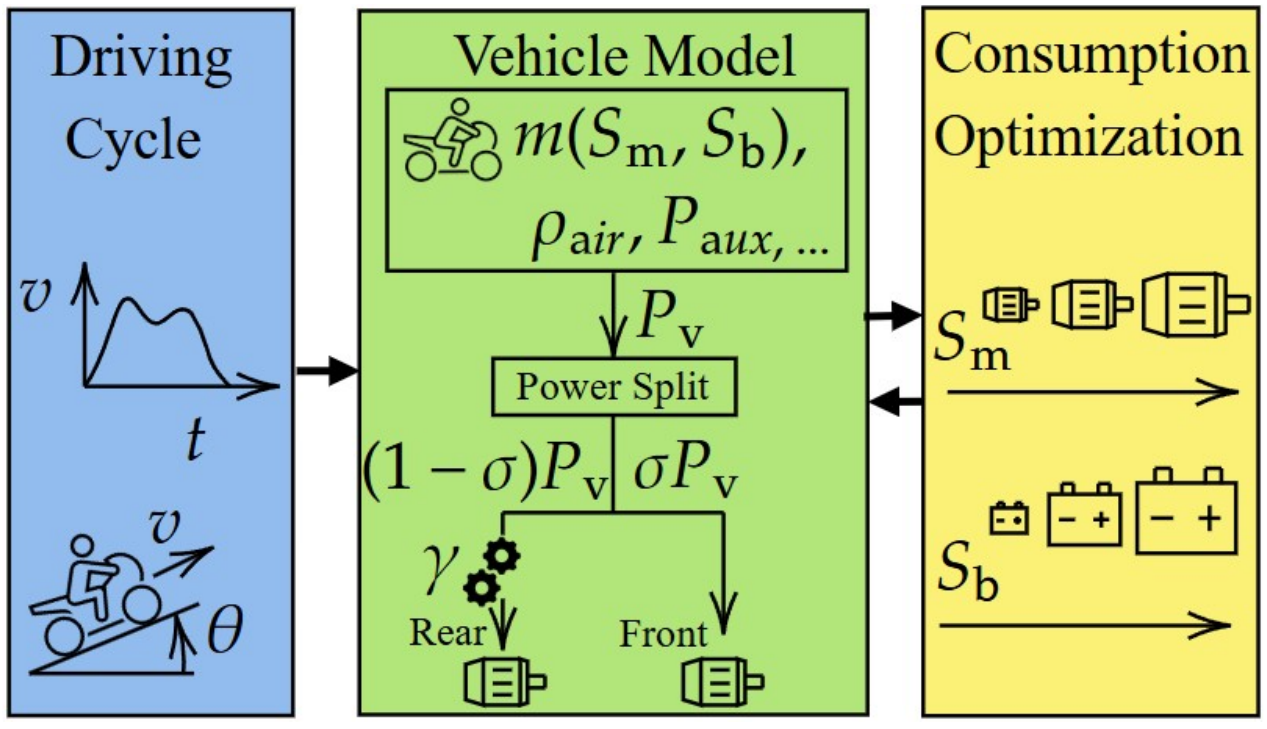}
    \caption{Schematic representation of our powertrain design framework. The driving cycle provides the vehicle kinematics for the vehicle operations. The power required and the force distribution are jointly optimized with the gear ratio, motor, and battery sizing, minimizing the energy consumption.}
    \label{fig: introduction}
\end{figure}

\vspace{-10pt}
According to the International Energy Agency~\citep{AR6}, 23\% of the global greenhouse gas emissions can be attributed to transportation.
In the collective effort to reduce the environmental impact of mobility, the scientific community and industry have developed alternatives to accelerate the transition to sustainable transportation.
On the one hand, the transition to battery electric vehicles can bring substantial environmental benefits when compared to conventional internal combustion vehicles.
On the other hand, these new technologies carry different challenges, such as the limited driving range and increased weight due to the relatively low battery energy density~\citep{trzesniowski2023powertrain} compared to fossil fuels. 
Hence, it becomes crucial to design the electric powertrain by jointly optimizing the sizes and operations of its components, minimizing mass and energy consumption, and effectively extending the vehicle's range.
In electric motorbikes, this strategy is particularly effective as the powertrain is particularly heavy compared to the vehicle's total weight, over 40\% of the total weight. 
This paper presents the powertrain design framework of Figure~\ref{fig: introduction} to optimize the gear ratio, rear motor size, battery size, and force distribution between the front and rear motors of an electric superbike.
To this end, we instantiate the energy consumption minimization problem, while accounting for adherence limits and performance constraints.

\textit{Related Literature}: The optimal design of an electric motorbike combines the challenges of traditional two-wheeler engineering with those of electric powertrain design.
One of the most significant advantages of battery electric motorcycles is the simplification of the power transmission, allowing for a fixed-gear transmission and keeping an immediate power delivery without clutch or shifts, thanks to the versatility of the electric motor.
Furthermore, an electric motor can be run at higher speeds, reaching higher power densities and reducing the motor mass and volumes~\citep{pyrhonen2009design}.
For this reason, optimizing the size and operation of the electric motors is crucial regarding energy consumption and performance.
While a downsized motor exhibits lower efficiency at high power and suffers from thermal issues, an oversized motor has low efficiency at low power, leading to unnecessary mass- and cost increments~\citep{pyrhonen2009design,AC}.
Likewise, the optimization of the gear ratio becomes critical to adequately match the vehicle and the motor speed~\citep{husain2021electric}, operate in the most efficient region of the map~\citep{su13137285}, and achieve top speed and maximum acceleration performance~\citep{guzzella2007vehicle}.
Moreover, the high flexibility of an electric drive, only necessitating a cable connection to deliver power, easily accommodates a two-wheel-driven architecture featuring a front in-wheel hub motor.
This way, it is possible to leverage the full potential of regenerative braking to minimize energy consumption ~\citep{Gao1999InvestigationOT}, without introducing further complexity and costs. 
Furthermore, an all-driven vehicle architecture improves handling and riding safety~\citep{Towards,Lot}, optimal torque repartition, and its influence on energy consumption~\citep{doi:10.1080/00423114.2024.2335261,Niccolai_2024}.
Finally, while the battery supplies energy to the powertrain, it remains the heaviest component of the entire system, providing a trade-off in terms of energy consumption and driving range ~\citep{su16177529}.
The joint sizing of the electric motor and the battery was analyzed by \cite{ClementeSalazarEtAl2022,ClementeSalazarEtAl2024}, where the authors propose a convex powertrain design framework to reduce energy consumption and costs for a family of vehicles.
To the best of the authors' knowledge, the joint optimization of the electric motor, gear ratio, force distribution, and battery for an electric superbike powertrain has been overlooked.

\textit{Contribution:} In this paper, we present a powertrain optimization framework to design the rear motor, gear ratio, and battery of a two-wheel-driven motorcycle.
Specifically, we account for the force distribution between the wheels and the tire adherence limits due to the harsh dynamic conditions typical of the operations.

\textit{Organization:} The remainder of this study is structured as follows: In Section~\ref{sec:methodology} we instantiate our framework, presenting the optimization variables, detailed components' models and the problem formulation, while Section~\ref{sec:results} showcases the application of our framework on a benchmark design problem.
The conclusions of the study are summarized in Section~\ref{sec:conclusions}, together with an outline of future research.

\section{Methodology}\label{sec:methodology}

In this section we present our powertrain design optimization framework, accurately describing all the components models and the constraints of the optimization problem.
Section~\ref{subsec:longitudinaldynamics} introduces the superbike kinematic, while Section~\ref{subsec: Masses} describes the influence of the components on the overall mass of the vehicle.
Section~\ref{subsec: electric machines} details the electric machines models and Section~\ref{subsec:optimizationvariables} explains the force distribution strategy.
In Section~\ref{subsec: battery} we show the inverter and battery models, giving an overview of the performance constraints included in Section~\ref{subsec: performance constraints}, and introducing energy recuperation indicators in Section~\ref{subsec:RRAE}.
Finally, we present the objective function and problem formulation in Section~\ref{subsec:formulation}, discussing the assumptions and limitations of our work in Section~\ref{subsec:discussion}.


\subsection{Longitudinal Dynamics}\label{subsec:longitudinaldynamics}

The force required at the wheel is based on driving cycles that provide the vehicle longitudinal speed $v$, acceleration $a$, and road inclination $\theta$.
The one degree of freedom model maps from the vehicle kinematics to the vehicle dynamics, neglecting tire slip and suspension dynamics.
Furthermore, the power required to follow the driving cycle depends on the vehicle's mass, which, in turn, is influenced by the components' sizing.
Under these hypotheses, the power required at the wheels can be written as 
\begin{equation}\label{eq:Pv}
     P_{\mathrm{v}} = m \cdot v \cdot \left( c_{\mathrm{r}} \cdot g \cdot \cos(\theta) + g \cdot \sin(\theta) + a \right) + \frac{1}{2} \cdot \rho \cdot C_{\mathrm{d}} \cdot A_{\mathrm{f}} \cdot v^3,
 \end{equation}
where $m$ is the overall mass, $g$ is the gravity acceleration constant, $c_{\mathrm{r}}$ is the tire rolling resistance coefficient, $\rho$ is the air density, $C_{\mathrm{d}}$ is the longitudinal aerodynamic drag coefficient, and $A_{\mathrm{f}}$ is the front cross-section area.

\subsection{Mass}\label{subsec: Masses}

The overall mass depends on the rear motor and battery sizing, with the addition of constant terms, such as the rider mass $m_{\mathrm{r}}$, and the glider mass $m_\mathrm{0}$ (including the front motor)
\begin{equation}
    m = m_\mathrm{0} + m_\mathrm{r} + \overline{m}_\mathrm{b} \cdot S_\mathrm{b} + \overline{m}_\mathrm{m}^\mathrm{r} \cdot S_\mathrm{m}^\mathrm{r},
    \label{eqn: overall mass}
\end{equation}
where $\overline{m}_\mathrm{b}$ and $\overline{m}_\mathrm{m}^\mathrm{r}$ are the reference battery and rear motor mass, respectively. The scaling factors for the battery $S_\mathrm{b}$ and for the rear motor $S_\mathrm{m}^\mathrm{r}$ linearly adjust the size of the components to satisfy the power demand.

\subsection{Electric Machines}\label{subsec: electric machines} 

The front in-wheel motor, characterized in the work of \cite{doi:10.1080/00423114.2024.2335261}, is directly coupled to the wheel without any reduction.
The power required for the motor $P_\mathrm{m}$ can be derived from the motor force at the wheel $F_\mathrm{m}$ and the vehicle speed.
Conversely, the rear frame-mounted motor is connected to the wheel via a single-speed fixed-gear transmission.
For this reason, we also need to account for the driveline efficiency $\eta_\mathrm{gb}$ when estimating the power required $P_\mathrm{m}^\mathrm{r}$
\begin{equation}
    \label{eqn: front power}
    P_\mathrm{m}^\mathrm{f} = F_\mathrm{m}^\mathrm{f} \cdot v
\end{equation}
\begin{equation}
    \label{eqn: rear power}
    P_\mathrm{m}^\mathrm{r} = 
    \begin{cases}
           \frac{1}{\eta_\mathrm{gb}} F_\mathrm{m}^\mathrm{r} \cdot v \quad & \text{if } F_{\mathrm{m}}^\mathrm{r}  \geq 0 \\
            F_\mathrm{m}^\mathrm{r} \cdot v \cdot \eta_\mathrm{gb} \quad & \text{if } F_{\mathrm{m}}^\mathrm{r} < 0
    \end{cases},
\end{equation}
where the superscripts $\mathrm{f}$ and $\mathrm{r}$ indicate the front and rear wheels, respectively.
Furthermore, the motor speeds $\omega_\mathrm{m}^\mathrm{f}$ and $\omega_\mathrm{m}^\mathrm{r}$ can be computed directly, assuming null tire slip, as 
\begin{equation}
    \omega_\mathrm{m}^\mathrm{f} = \frac{v}{r_\mathrm{w}^\mathrm{f}},
\end{equation}
\begin{equation}
    \omega_\mathrm{m}^\mathrm{r} = \frac{v \cdot \gamma}{r_\mathrm{w}^\mathrm{r}}.
\end{equation}
where $r_\mathrm{w}^\mathrm{f}$ and $r_\mathrm{w}^\mathrm{r}$ are the wheels radii and $\gamma$ the gear ratio of the transmission.
We can constrain the maximum value of $\gamma$ linking the maximum speed reachable by the rear motor $\omega_\mathrm{m,max}^\mathrm{r}$ to the maximum vehicle speed $v_\mathrm{max}$
\begin{equation}
\gamma \leq \frac{\omega_{\mathrm{m,max}}^\mathrm{r} \cdot r_\mathrm{w}^r}{v_\mathrm{max}}.
    \label{eqn: up bound gamma}
\end{equation}
In our framework we consider a front motor of constant size, while linearly scaling the rear motor along its axial dimension, varying its power $P_\mathrm{m,max}^\mathrm{r}$ and torque $T_\mathrm{m,max}^\mathrm{r}$ with respect to the reference motor  
\begin{equation}
    S_\mathrm{m}^\mathrm{r} =  \frac{P_\mathrm{m,max}^\mathrm{r}}{\overline{P}_{\mathrm{m,max}}^\mathrm{r}} = \frac{T_\mathrm{m,max}^\mathrm{r}}{\overline{T}_\mathrm{m,max}^\mathrm{r}},
    \label{eq:Sm}
\end{equation}
where $S_\mathrm{m}^\mathrm{r}$ is the rear electric machine scaling factor, $\overline{P}_{\mathrm{m,max}}^\mathrm{r}$ and $\overline{T}_\mathrm{m,max}^\mathrm{r}$ are the reference motor maximum power and torque, respectively.
\begin{figure}[t]
    \centering
    \includegraphics[width=\columnwidth]{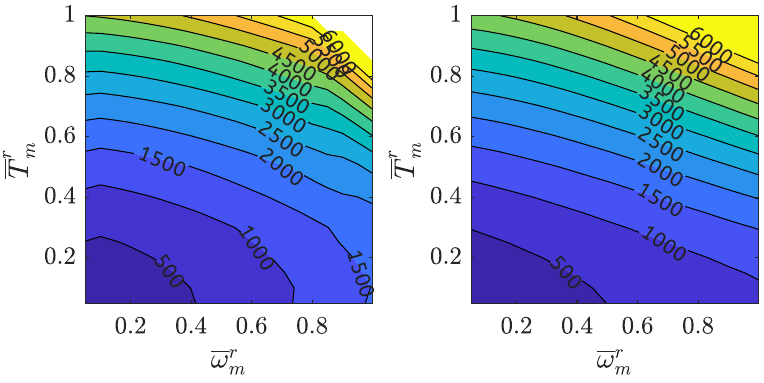}
    \caption{Comparison between the normalized data (left) of the rear electric motor power losses (in W) and our model (right). The $R^2$ was equal to 0.99, with an NMRSE of 2.8\%.}
    \label{fig:EMe}
\end{figure}
While for the front motor losses we used the same approach as adopted by \cite{doi:10.1080/00423114.2024.2335261}, the rear motor map is directly provided by Ducati, and its power losses are estimated using a polynomial fit of the speed and motor torque (Fig. \ref{fig:EMe}), described in \cite{7903710}, and scaling the iron, mechanical, and copper losses of the reference motor, $\overline{P}_\mathrm{Fe}^\mathrm{r}$, $\overline{P}_\mathrm{Mech}^\mathrm{r}$, and $\overline{P}_\mathrm{Cu}^\mathrm{r}$ as
\begin{equation}
 P_\mathrm{Fe}^\mathrm{r} = S_\mathrm{m} \cdot a_\mathrm{Fe} \cdot \omega_\mathrm{m}^\mathrm{r},
\end{equation} 
\begin{equation}
\resizebox{0.89\hsize}{!}{$
 P_\mathrm{Mech}^\mathrm{r} = S_\mathrm{m}\left(a_\mathrm{Mech} + b_\mathrm{Mech} \cdot \omega_\mathrm{m}^\mathrm{r} + c_\mathrm{Mech} (\omega_\mathrm{m}^\mathrm{r})^2 + d_\mathrm{Mech}(\omega_\mathrm{m}^\mathrm{r})^3 \right)$,}
\end{equation}
\begin{equation}
    \label{eq: copper losses}
\resizebox{0.88\hsize}{!}{$
     P_\mathrm{Cu}^\mathrm{r} = \left(\frac{S_\mathrm{m} \cdot l_\mathrm{co}}{ l_\mathrm{co}+l_\mathrm{ew}} + \frac{l_\mathrm{ew}}{ l_\mathrm{co}+l_\mathrm{ew}}\right) \left( a_\mathrm{Cu} + b_\mathrm{Cu} \left( \omega_\mathrm{m}^\mathrm{r}\right)^2 \right)  \left(\frac{P_\mathrm{m}^\mathrm{r}}{\omega_\mathrm{m}^\mathrm{r}S_\mathrm{m}}\right)^2$,}
\end{equation}
where $a_\mathrm{F_e}$, $a_\mathrm{Mech}$, $b_\mathrm{Mech}$, $c_\mathrm{Mech}$, $d_\mathrm{Mech}$, $a_\mathrm{Cu}$, and $b_\mathrm{Cu}$ are identified loss coefficients, $l_\mathrm{co}$ is the stack length, and $l_\mathrm{ew}$ is the end-winding length.
Each motor is also bound to not exceed its operational limits
\begin{equation}\label{eq: max front power}
    P_\mathrm{m}^\mathrm{f} \leq P_\mathrm{m,max}^\mathrm{f},
\end{equation}
\begin{equation}\label{eq: max rear power}
    P_\mathrm{m}^\mathrm{r} \leq \overline{P}_\mathrm{m,max}^\mathrm{r} \cdot S_\mathrm{m},
\end{equation}    
 \begin{equation}\label{eq: max front torque}
    F_\mathrm{m}^\mathrm{f} \cdot r_\mathrm{w}^\mathrm{f} \leq T_\mathrm{m,max}^\mathrm{f},
\end{equation}
\begin{equation}\label{eq: max rear torque}
    \frac{P_\mathrm{m}^\mathrm{r}}{\omega_\mathrm{m}^\mathrm{r}}  \leq \overline{T}_\mathrm{m,max}^\mathrm{r} \cdot S_\mathrm{m},
\end{equation}
where $\overline{P}_\mathrm{m,max}^\mathrm{r}$ is the maximum reference rear motor power, $ T_\mathrm{m,max}^\mathrm{f}$ the maximum front motor torque, and $\overline{T}_\mathrm{m,max}^\mathrm{r}$ the maximum rear reference motor torque.

Finally, we can write the overall electric motor power $P_\mathrm{ac}$ as
\begin{equation}
  P_\mathrm{ac} = P_\mathrm{m}^\mathrm{f} + P_\mathrm{loss}^\mathrm{f}  + P_\mathrm{m}^\mathrm{r} + P_\mathrm{Cu}^\mathrm{r} + P_\mathrm{Fe}^\mathrm{r} +  P_\mathrm{Mech}^\mathrm{r}
\end{equation}

\subsection{Propulsion Blending}\label{subsec:optimizationvariables}  

High-performance motorcycle riding is characterized by high load transfers from one wheel to the other.
Hence, keeping the same braking effort on a wheel independently of the load transfer might result in locking, with serious consequences on safety. 
For this reason, the power split between the front and rear wheels $\sigma$ might be based on the strategy that maximizes tire adherence $\sigma^\mathrm{a}$ to ensure vehicle stability~\citep{doi:10.1080/00423114.2024.2335261}
 \begin{equation}
         \sigma^\mathrm{a} = \frac{b}{w_\mathrm{b}} - \frac{h}{w_\mathrm{b}}\left(\frac{X_\mathrm{f} + X_\mathrm{r}}{m \, g \, cos(\theta)} - c_\mathrm{r}\right),
         \label{eq:ideal braking bias}
 \end{equation}
where $X_\mathrm{f}$ and $X_\mathrm{r}$ are the longitudinal front and rear wheel force, respectively, $b$ is the longitudinal distance between the rear wheel contact point to the vehicle center of mass, $w_\mathrm{b}$ is the distance between the front and rear axles (wheelbase), and $h$ is the height of the vehicle center of mass.
Since this strategy is not necessarily the optimal choice when minimizing the energy consumption, to ensure the motor is operated in the most efficient region we jointly optimize the gear ratio $\gamma$ and $\sigma$, distributing the longitudinal power required $P_{\mathrm{v}}$ between front $P_{\mathrm{v}}^\mathrm{f}$ and rear $P_{\mathrm{v}}^\mathrm{r}$ as
 \begin{equation}
     P_{\mathrm{v}}^\mathrm{f} = \sigma {P_{\mathrm{v}}},
     \label{eqn: front required power}
 \end{equation}
 \begin{equation}
     P_{\mathrm{v}}^\mathrm{r} = (1-\sigma){P_{\mathrm{v}}},
     \label{eqn: rear required power}
 \end{equation}
  \begin{equation}
     \sigma \geq 0
     \label{eqn: power split lower}
 \end{equation}
   \begin{equation}
     \sigma \leq 1
     \label{eqn: power split upper}
 \end{equation}
However, we still account for safety constraints by avoiding the locking of the front wheel
 \begin{equation}
     \sigma \leq \sigma^\mathrm{a}.
     \label{eqn: power split adherence}
 \end{equation}
The net force acting on each wheel is further divided among the force supplied by the motors $F_{\mathrm{m}}^\mathrm{f}$ and $F_{\mathrm{m}}^\mathrm{r}$, the dissipative brakes $F_{\mathrm{brk}}^\mathrm{f}$ and $F_{\mathrm{brk}}^\mathrm{r}$, and the tractive force $F_{\mathrm{tr}}^\mathrm{f}$.
The tractive term accounts for the force transfer from front wheel to the other in the case of motor saturation, redistributing the missing quote on the other wheel. 
While the front wheel locking during braking is prevented by equation \eqref{eqn: power split adherence}, we impose a rear tire threshold $\mu_\mathrm{brk,peak}^\mathrm{r}$ constraint on the rear wheel adherence $\mu^\mathrm{r}$
\begin{equation}
    \mu^\mathrm{r} = \frac{X_\mathrm{r}}{\frac{m \cdot g \cdot \cos(\theta)}{w_\mathrm{b}}(w_\mathrm{b}-b- h \cdot c_\mathrm{r}) + \frac{h}{w_\mathrm{b}}\left(X_\mathrm{f} + X_\mathrm{r}\right)},
    \label{eq: rear adherence limit}
\end{equation}
\begin{equation}
 \mu^\mathrm{r} \geq \mu_\mathrm{brk,peak}^\mathrm{r},
    \label{eq: rear adherence limit max}
\end{equation}
so that the rear tire adherence is greater than or equal (braking adherence has negative values) to the maximum adherence allowed to prevent the wheel locking.
Furthermore, the required braking force is applied in a serial fashion: first, by regenerative braking, then by mechanical braking
\begin{equation}
\label{eqn:front req force}
    F_{\mathrm{v}}^\mathrm{f} = \sigma \frac{P_{\mathrm{v}}}{v} = F_{\mathrm{m}}^\mathrm{f} + F_{\mathrm{brk}}^f + F_{\mathrm{tr}}^\mathrm{f},
\end{equation}
\begin{equation}
\label{eqn:rear req force}
F_{\mathrm{v}}^\mathrm{r} = (1-\sigma)\frac{P_{\mathrm{v}}}{\mathrm{v}} + F_{\mathrm{tr}}^\mathrm{f} = F_{\mathrm{m}}^\mathrm{r} + F_{\mathrm{brk}}^\mathrm{r}.
\end{equation}
Finally, we guarantee the mutual exclusion between the motor tractive and braking forces by introducing constraints on the sign of the repartitioned forces, ensuring coherence between the tractive and braking forces
 \begin{equation}
     F_{\mathrm{m}}^\mathrm{f} \; F_{\mathrm{m}}^\mathrm{r} \geq 0,
     \label{eqn: motor force coherence}
 \end{equation}
  \begin{equation}
     F_{\mathrm{brk}}^\mathrm{f} \leq 0,
     \label{eqn: front braking coherence}
 \end{equation}
 \begin{equation}
     F_{\mathrm{brk}}^\mathrm{r} \leq 0,
     \label{eqn: rear braking coherence}
 \end{equation}
  \begin{equation}
     F_{\mathrm{tr}}^\mathrm{f} \geq 0,
     \label{eqn: front traction coherence}
 \end{equation}
 \begin{equation}
     F_{\mathrm{m}}^\mathrm{f} F_{\mathrm{brk}}^\mathrm{f} \geq 0,
     \label{eqn: front braking and motor coherence}
 \end{equation}
 \begin{equation}
     F_{\mathrm{m}}^\mathrm{r} F_{\mathrm{brk}}^\mathrm{r} \geq 0.
     \label{eqn: rear braking and motor coherence}
 \end{equation}

\subsection{Battery}\label{subsec: battery}

The battery power $P_{\mathrm{b}}$ is equal to the sum of overall motor input power $P_{\mathrm{ac}}$ and the auxiliaries consumption $P_{\mathrm{aux}}$, considering the inverter and battery efficiency $\eta_\mathrm{inv}$ and $\eta_\mathrm{b}$
	\begin{align}
		\label{eq:inverter}
		P_{\mathrm{b}} = 
		\begin{cases}
			\left(\frac{ P_{\mathrm{ac}}}{\eta_\mathrm{inv}} + P_{\mathrm{aux}}\right)(\eta_\mathrm{b})^{-1} \quad & \text{if } P_{b} \geq 0 \\
			 \left(\eta_\mathrm{inv}P_{\mathrm{ac}} + P_{\mathrm{aux}}\right)\eta_\mathrm{b}\quad & \text{if } P_{\mathrm{b}} < 0
		\end{cases}.
	\end{align}
The power flow causes a change in battery energy $E_{\mathrm{b}}$, following the equation
\begin{equation}\label{eq:diff}
    \frac{\mathrm{d} E_{\mathrm{b}}}{\mathrm{d} t} = - 	P_{\mathrm{b}},
\end{equation}
leading to an overall energy consumption $E_{\mathrm{c}}$ computed as
 \begin{equation*}
    E_{\mathrm{c}}= \left(E_{\mathrm{b}}(0) - E_{\mathrm{b}}(T) \right),
 \end{equation*}
where $E_{\mathrm{b}}(0)$ is the energy at the beginning of the driving cycle and $E_{\mathrm{b}}(T)$ the energy remaining at its end.

We optimize the size of the battery by introducing a battery scaling factor $S_\mathrm{b}$, linking the full battery capacity $E_\mathrm{b,max}$ to the reference full battery capacity $\overline{E}_\mathrm{b,max}$
\begin{equation}
    S_\mathrm{b} = \frac{E_\mathrm{b,max}}{\overline{E}_\mathrm{b,max}}.
\label{eq:Sb}
\end{equation}
Ultimately, the battery energy is constrained in its operational range by 
\begin{equation}
    E_{\mathrm{b}} \in \left[\overline{E}_\mathrm{b,max} \cdot \xi_\mathrm{min}, \overline{E}_\mathrm{b,max} \cdot \xi_\mathrm{max}\right] \cdot S_\mathrm{b},
    \label{eq:Eblimits}
\end{equation}
through the coefficients of minimum and maximum state of charge $\xi_\mathrm{min}$ and $\xi_\mathrm{max}$.

\subsection{Performance Constraints}\label{subsec: performance constraints}

The driving cycle provides useful insights into the riding style, but it is not fully representative of all the performance required to rank the design motorcycle among the other vehicles present in the market. For instance, it does not constrain the riding range.
For this reason, we include performance constraints on maximum acceleration time, maximum speed, power gradeability, torque gradeability, and riding range  
\begin{equation}
    \label{eq:accT}
    \resizebox{\columnwidth}{!}{%
    $
    \begin{split}
 t_{\mathrm{a}} \geq \frac{m \cdot \omega_\mathrm{r}^\mathrm{f} \cdot r_{\mathrm{w}}^\mathrm{f}}{\frac{T_{\mathrm{m,max}}^\mathrm{f}}{r_\mathrm{w}^\mathrm{f}}+\frac{S_\mathrm{m} \cdot \gamma \cdot \overline{T}_{\mathrm{m,max}}^\mathrm{r} \cdot \eta_\mathrm{gb}}{r_\mathrm{w}^\mathrm{r}}} + \frac{m \cdot r_\mathrm{w}^\mathrm{r}}{ S_\mathrm{m} \cdot \gamma \cdot \overline{T}_{\mathrm{m,max}}^\mathrm{r} \cdot \eta_\mathrm{gb}}\Biggl[  (v_{\mathrm{f}} + \\
    -\omega_\mathrm{r}^\mathrm{f} \cdot r_\mathrm{w}^\mathrm{f}) - \frac{P_\mathrm{{m,max}}^\mathrm{f} \cdot  r_\mathrm{w}^\mathrm{r}}{\gamma  \cdot S_\mathrm{m} \cdot  \overline{T}_{\mathrm{m,max}}^\mathrm{r} \cdot \eta_\mathrm{gb}}\log\left(\frac{P_\mathrm{{m,max}}^\mathrm{f} + \frac{v_\mathrm{f} r_\mathrm{w}^\mathrm{r}}{ S_\mathrm{m}  \cdot \gamma \cdot \overline{T}_{\mathrm{m,max}}^\mathrm{r} \cdot \eta_\mathrm{gb}}}{P_\mathrm{{m,max}}^\mathrm{f} + \frac{\omega_\mathrm{r}^\mathrm{f} \cdot r_\mathrm{w}^\mathrm{f} r_\mathrm{w}^\mathrm{r}}{ S_\mathrm{m}  \cdot \gamma \cdot \overline{T}_{\mathrm{m,max}}^\mathrm{r} \cdot \eta_\mathrm{gb}}}\right) \Biggr],
    \end{split} 
    $
}
\end{equation}
\begin{equation}\label{eq:Vmax}
    P_\mathrm{m,max}^\mathrm{f} + S_{\mathrm{m}} \cdot \overline{P}_{\mathrm{m,max}}^\mathrm{r}\cdot\eta_\mathrm{gb} \geq \frac{1}{2}\rho \cdot C_{\mathrm{D}}^\mathrm{p} \cdot A_{\mathrm{f}}^\mathrm{p} \cdot v_{\mathrm{max}}^3 + m\cdot g \cdot c_\mathrm{r} \cdot v_{\mathrm{max}},
\end{equation}
\begin{equation}\label{eq:Pgrad}
    P_\mathrm{m,max}^\mathrm{f} + S_{\mathrm{m}} \cdot \overline{P}_{\mathrm{m,max}}^\mathrm{r} \cdot \eta_\mathrm{gb} \geq m \cdot g \cdot \sin(\theta_\mathrm{max}) \cdot v_{\mathrm{min}},
\end{equation}
\begin{equation}\label{eq:Tgrad}
    \frac{T_\mathrm{m,max}^\mathrm{f}}{r_\mathrm{w}^\mathrm{f}}+\frac{S_{\mathrm{m}} \cdot \overline{T}_{\mathrm{m,max}}^\mathrm{r} \cdot \gamma}{r_\mathrm{w}^\mathrm{r}} \geq m \cdot g \cdot \sin(\theta_\mathrm{max}),
\end{equation}
\begin{equation}\label{eq:Range}
    E_{\mathrm{b}}(0) - E_{\mathrm{b}}(T) \leq S_{\mathrm{b}} \cdot \left(\xi_\mathrm{max} - \xi_\mathrm{min} \right) \overline{E}_\mathrm{b,max}\frac{D_\mathrm{c}}{D_\mathrm{r}},
\end{equation}
where $t_{\mathrm{a}}$ is the maximum acceleration time on flat road from $0$ to $v_{\mathrm{f}}$, $v_{\mathrm{max}}$ is the top speed, $D_\mathrm{r}$ the minimum riding range, $D_\mathrm{c}$ the cycle length, $v_{\mathrm{min}}$ is the speed at which the vehicle can climb a road inclined by the angle $\theta_\mathrm{max}$. The superscript $\mathrm{p}$ indicates a rider sitting prone on the motorcycle, for a 10\% air resistance power reduction~\citep{cossalter2006motorcycle}. 
Finally, $t_\mathrm{a}$ considers the front motor operating in maximum torque until the rate speed is reached ($\omega_\mathrm{r}^\mathrm{f}$) and then in the maximum power region, while the rear is constantly in the maximum torque region, a typical condition for the reference speed $v_\mathrm{f}$ of 100 \unit{km/h}.

\subsection{Regenerative Ratio and Average Efficiency}\label{subsec:RRAE}

We introduce the regenerative ratio $\zeta$ and the average efficiency $\eta$ as a measure to understand the benefits of the propulsive blending strategy and its influence on energy consumption:
\begin{equation*}
    \zeta = \frac{|E_\mathrm{b,in}|}{E_\mathrm{b,out}}\cdot100,
\end{equation*}
\begin{equation*}
    \eta = \frac{E_\mathrm{v,tr} + E_\mathrm{v,brk}}{E_\mathrm{b,out} + E_\mathrm{b,in}}\cdot 100 = \frac{E_\mathrm{v,tr} + E_\mathrm{v,brk}}{E_\mathrm{b,out}(1- \zeta/100)}\cdot 100,
\end{equation*}
where $E_\mathrm{v,tr}$ is the vehicle or mechanical traction energy (positive), $E_\mathrm{v,brk}$ is the vehicle braking energy (negative), $E_\mathrm{b,out}$ is the overall energy flowing out of the battery (positive), and $E_\mathrm{b,in}$ flowing in (negative). 
While both indicators can vary from 0 to 100~\% (ideal powertrain), $\zeta$ indicates how much energy is recuperated during the drive cycle in absolute terms, while $\eta$ takes into account the energy needed for the vehicle propulsion as well.
They link the energy reduction resulting from the joint design to the specific contributions of net lower mechanical energy (numerator) and higher regeneration ratio.

\begin{table*}[t!]
    \caption{Output of the powertrain design optimization for different driving cycles. We compare the adherence-driven propulsive strategy $\sigma^\mathrm{a}$ with the optimum found by our framework $\sigma^\star$.}
    \centering
    \begin{tabular}{|c|cccc|cccc|}
        \hline
        \multirow{2}{*}{\textbf{Cycle}} & \multicolumn{4}{c|}{$\sigma^{a}$} & \multicolumn{4}{c|}{$\sigma^\star$} \\
         & $m_\mathrm{v}$  & $E_\mathrm{b,max}$  & $P_\mathrm{m,max}^\mathrm{r}$  & $\bar{E_\mathrm{c}}$ & $m_\mathrm{v}$& $E_\mathrm{b,max}$  & $P_\mathrm{m,max}^\mathrm{r}$ & $\bar{E_\mathrm{c}}$ \\
        \ & (\SI{}{\kilo\gram}) & (\SI{}{\kilo\watt\hour}) & (\SI{}{\kilo\watt}) & (\SI{}{\frac{\watt\hour}{\kilo\meter}}) & (\SI{}{\kilo\gram})  & (\SI{}{\kilo\watt\hour}) & (\SI{}{\kilo\watt}) & (\SI{}{\frac{\watt\hour}{\kilo\meter}})
        \\
        \hline
         WMTC 1 & 167.95 & 5.39 & 78.57 & 43.12 & 159.63 (-4.95\%) & 4.29 (-20.50\%) & 75.62 (-3.75\%) & 34.28 (-20.50\%)  \\
        \hline
         WMTC 2 & 175.12 & 6.34 & 81.11 & 50.75  & 170.64 (-2.56\%) & 5.75 (-9.4\%) & 79.52 (-1.96\%) & 45.97 (-9.4\%)   \\
        \hline
         WMTC 3 & 204.76 & 10.28 & 91.62 & 82.23 &  201.13 (-1.77\%) & 9.80 (-4.68\%) & 90.33 (-1.40\%) & 78.38 (-4.68\%)  \\
        \hline
         WMTC & 190.67 & 8.41 & 86.64 & 67.3  & 185.85 (-2.54\%) & 7.77 (-7.65\%) & 84.92 (-1.98\%) & 62.15 (-7.65\%)   \\
        \hline
         UDDS & 166.00 & 5.13 & 77.88 & 41.04  & 159.44 (-3.94\%) & 4.26 (-16.96\%) & 75.56 (-2.98\%) & 34.07 (-16.96\%)   \\
        \hline
         Sport & 224.71 & 12.93 & 98.69 & 103.45  & 202.95(-9.69\%) & 10.04 (-22.36\%) & 90.98 (-7.82\%) & 80.31 (-22.36\%)  \\
        \hline
    \end{tabular}
    \label{tab:doubleAct}
\end{table*}

\subsection{Objective Function and Problem Formulation}\label{subsec:formulation}

We formulate the powertrain design optimization by minimizing the objective function $J$ 
\begin{equation}
     J = E_{\mathrm{c}} + w\sum_{t = 1}^T  F_{\mathrm{tr}}^\mathrm{f} ,
\end{equation}
composed of the energy consumption over the driving cycle and tractive terms responsible for the force transfer in the case of motor saturation, the weighted via the parameter $w$.
In fact, including the last term is essential to enforce the force repartition between the motors, depending on the bias strategy.

\begin{problem}[\textbf{Powertrain Design}]\label{prb:CPDP}
The optimal powertrain design and propulsion blending strategy to minimize energy consumption are the solution to the joint design optimization problem
\begin{align*}
    & \qquad \min_{\gamma, \sigma, S_\mathrm{m}, S_\mathrm{b}} \qquad J\\
    & \text{s.t. } \; \text{Powertrain Constraints \eqref{eq:Pv}-\eqref{eq:Eblimits}}\\
    & \quad \;\;\;\; \text{Performance Constraints \eqref{eq:accT}-\eqref{eq:Range}}
\end{align*}
\end{problem}
Therefore, we can compute the optimal solution with standard nonlinear programming methods.

\vspace{-8pt}
\subsection{Discussion}\label{subsec:discussion}

In this section, we present some clarification on the assumptions and limitations of our work.
First, we assumed a linear scaling of the rear electric machine mass, torque, power, and losses, according to the conventional axial scaling laws for synchronous permanent magnet motors, while keeping the dimension of the front motor constant. 
Second, we considered changes in gear ratio to not significantly affect vehicle mass and driveline efficiency.
Likewise, we neglected the effect of displacements of the center of gravity owing to component sizing, and we considered the aerodynamic forces acting on the vehicle's center of mass.
Despite these simplifications, the propulsive blending optimization is still valid,  as it is independent of vehicle geometry and only dependent on motor losses \citep{doi:10.1080/00423114.2024.2335261}.
Finally, we represented inverter and battery losses using constant efficiency values. 
Incorporating more sophisticated battery models would enable accounting for potential power limitations that could impact the solution, and comparing the results of the framework for various chemistries would provide useful insights for battery technology selection.
Although extending the approach to more advanced models is beyond the scope of this paper, it remains feasible, as the problem is non-convex and solved using general off-the-shelf solvers.

\section{Results}\label{sec:results}

\begin{table}
\caption{Vehicle and Performance Parameters.}
    \centering
\begin{tabular}{lccc}
     \toprule
     \textbf{Symbol} & \textbf{Value} & \textbf{Unit} & \textbf{Description} 
     \\
     \midrule
     $m_\mathrm{r}$ & 80 & \SI{}{\kilo\gram} & Rider mass\\
     $m_0$ & 75 & \SI{}{\kilo\gram} & Glider mass\\
     $r_\mathrm{w}^\mathrm{f}$ & 0.321 & \SI{}{\meter} & Front wheel radius \\
     $r_\mathrm{w}^\mathrm{r}$ & 0.318 & \SI{}{\meter} & Rear wheel radius \\
     $h$ & 0.573 &  \SI{}{\meter} & CoG vertical height \\
     $b$ & 0.6935 &  \SI{}{\meter} & CoG longitudinal distance \\
     $w_\mathrm{b}$ & 1.37 &  \SI{}{\meter} & Wheelbase\\
     $c_\mathrm{r}$ & 0.015 & - & Tire rolling resistance\\
     $C_\mathrm{d}A$ & 0.32 & \SI{}{\meter^2} & Drag coefficient\\
     $\mu_\mathrm{brk,max}$ & -0.8 & - & Braking rear adherence limit\\
     $\rho_\mathrm{air}$ & 1.25 & \SI{}{\frac{\kilo\gram}{\meter^3}} & Air density\\
     $g$ & 9.81 & \SI{}{\frac{\meter}{\second^2}} & Gravitational acceleration\\
     $\eta_\mathrm{gb}$ & 0.96 & - & Gearbox efficiency\\
     $\eta_\mathrm{b}$ & 0.92 & - & Battery efficiency\\
     $\eta_\mathrm{inv}$ & 0.96 & - & Inverter efficiency\\
     $P_\mathrm{aux}$ & 100 & \SI{}{\watt} & Auxiliary power\\
     $\xi_\mathrm{min}$ & 0.1 & - & Min battery SoC \\
     $\xi_\mathrm{max}$ & 0.9 & - & Max battery SoC \\
     $v_\mathrm{max}$ & 250 & \SI{}{\frac{\kilo\meter}{\hour}} & Top speed\\
     $\theta_\mathrm{max}$ & 25 & \% & Max road slope climbing\\
     $v_\mathrm{min}$ & 15 & \SI{}{\frac{\kilo\meter}{\hour}} & Min climbing speed\\
     $v_\mathrm{f}$ & 100 & \SI{}{\frac{\kilo\meter}{\hour}} & Final acceleration velocity \\
     $t_\mathrm{a}$ & 3.5 & \SI{}{\second} & Min acceleration time \\
     $D_\mathrm{r}$ & 100 & \SI{}{\kilo\meter} & Min riding range\\
     $w$ & $10^3$ & - & Cost function weight\\
     \bottomrule
     \label{tab: parameters}
\end{tabular}
\end{table}

\begin{table}
    \caption{Regenerative ratio ($\zeta$) and efficiency ($\eta$) for the adherence and optimized propulsive biases ($\sigma^\mathrm{a}$, $\sigma^\star$) and different driving cycles.}
    \centering
    \begin{tabular}{|c|cc|cc|}
        \hline
        \multirow{2}{*}{\textbf{Cycle}} & \multicolumn{2}{c|}{$\sigma^{a}$} & \multicolumn{2}{c|}{$\sigma^\star$} \\
         & $\zeta$ & $\eta$ & $\zeta$ & $\eta$ \\
        \ & (\%) & (\%) & (\%) & (\%)
        \\
        \hline
         WMTC 1 & 6.28 & 44.26 & 11.77 & 54.4 \\
        \hline
         WMTC 2 & 6.52 & 63.62 & 8.92 & 69.76 \\
        \hline
         WMTC 3 & 1.47 & 74.49 & 1.85 & 77.95 \\
        \hline
         WMTC & 3.22 & 68.74 & 4.61 & 74.07 \\
        \hline
         UDDS & 5.42 & 49.24 & 10.23 & 58.33\\
        \hline
         Sport &  14.60 & 47.78 & 24.40 & 59.64 \\
        \hline
    \end{tabular}
    \label{tab: average efficiency}
\end{table}
 
\begin{figure}[t]
	\centering
	\includegraphics[width=\columnwidth]{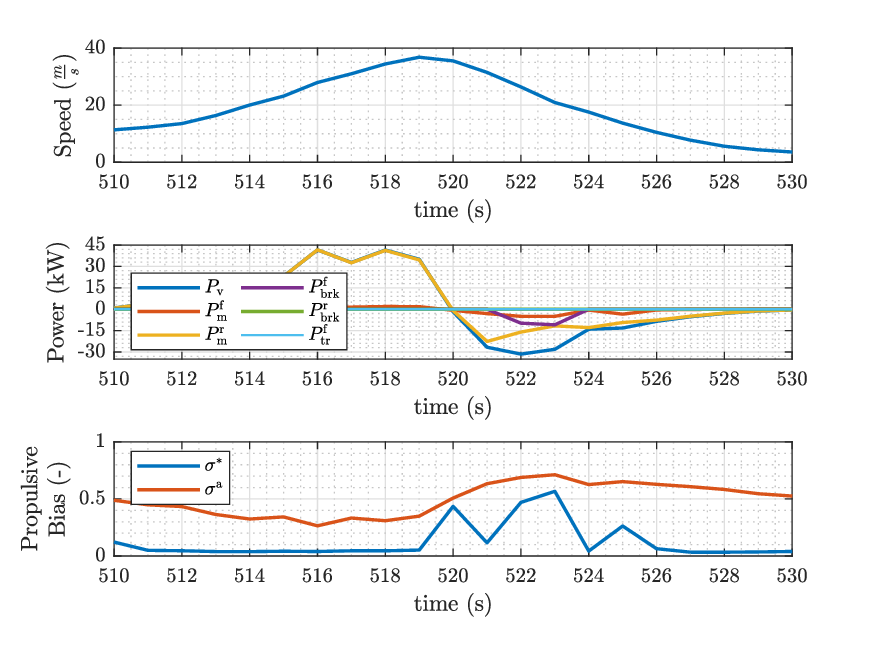}
	\caption{A 20-second window of the Sport drive cycle, showing the speed profile followed by the vehicle at the top, the power repartition with the optimal propulsive strategy $\sigma^\star$               in the middle, and the comparison between the two different strategies $\sigma^\mathrm{a}$ and $\sigma^\star$ at the bottom.}
	\label{fig: driving cycle result}
\end{figure}

To showcase the effectiveness of our approach, we applied our superbike design framework to different standard regulatory driving cycles (UDSS, WMTC) and custom cycles based on real-world data via a Monte Carlo Markov Chain approach (Sport).
In our case study, we compare the design of the adherence-driven propulsive blending strategy ($\sigma^\mathrm{a}$) with the optimal solution of our framework ($\sigma^\star$), with the vehicle and performance parameters of Table~\ref{tab: parameters}.
We discretize the problem with a sampling time of 1~\unit{s} and the forward Euler method, and we solve it using the nonlinear solver IPOPT~\citep{Wachter2006} after parsing it with CasADi~\citep{Andersson2019}.
The optimization problem involves a total of 41473 variables, with 16229 equality constraints and 73932 inequality constraints. The solver converged to an optimal solution in 441 iterations, which took approximately 491 seconds.
We compare the vehicle's mass, battery capacity, rear motor power, and specific energy consumption of both strategies in Table~\ref{tab:doubleAct}.
The results suggest that the joint optimization of sizing and force distribution can significantly reduce the mass and average energy consumption over the driving cycle $\bar{E_\mathrm{c}}$, allowing for smaller and cheaper motors and batteries.
In \emph{dynamic} cycles like  ``Sport", more energy is recovered by braking (higher $\zeta$ values, Table \ref{tab: average efficiency}) due to more frequent opportunities to regenerate energy. 
However, the highest efficiency (77.95\%) is still achieved in the \emph{smooth}  WMTC part 3 cycle because of the lower speed and acceleration profile.
In particular, Figure~\ref{fig: driving cycle result} shows a 20-second window of the ``Sport" cycle, underlining the difference between $\sigma^\mathrm{a}$ described in equation \eqref{eq:ideal braking bias} and the optimized propulsive bias $\sigma^\star$, displaying the braking repartition dynamic that maximizes the energy recuperation, while still accounting for driving safety by avoiding wheel locking.
Finally, the gear ratio is not displayed in the tables as it remains constant for all the cases, bounded to its maximum value to satisfy the performance constraint on the acceleration time, as any decrease would worsen the efficiency of the operating points in the trade-off between torque and motor speed.

\section{Conclusions}\label{sec:conclusions}

This paper presents a framework to optimize the design and operations of a two-wheel-driven electric superbike Powertrain.
Specifically, we optimize the gear ratio, force distribution (propulsion blending strategy), and rear motor and battery size to minimize energy consumption.
To this end, we devised the vehicle energy consumption model, including a scalable power loss model for the rear motor based on Ducati efficiency maps and accounting for performance constraints and adherence limits.
We applied our framework to a case study analyzing different driving cycles, comparing the adherence-driven blending strategy with our optimal solution.
The joint design of the propulsion strategy and components significantly reduced the vehicle's energy consumption compared to the sole optimization of the components with the adherence-driven blending strategy, ranging from 4.68\% (WMTC part 3) to 22.36\% (Sport).
This result highlights the importance of jointly optimizing the powertrain and the force distribution in electric motorcycles and demonstrates the capabilities of our framework.

Finally, this work opens the field for further development, such as developing a cost model, extending the size optimization to the front motor, and including constraints on the vehicle's lateral dynamics.

\begin{ack}
We wish to thank Ducati Motor Holding S.P.A for the motor data availability and Dr. I. New for proofreading
this paper. This publication is part of the project NEON with project number 17628 of the research program Crossover which is (partly) financed by the Dutch Research Council (NWO).
\end{ack}

\bibliography{Bibliography}             



\end{document}